\documentclass[aps,prl,twocolumn,showpacs]{revtex4}
\usepackage{graphicx}

\begin{document}

\title{Simultaneous Vanishing of the Nematic Electronic State and the Structural Orthorhombicity in NaFe$_{1-x}$Co$_x$As Single Crystals}

\author{Qiang Deng, Jianzhong Liu, Jie Xing, Huan Yang, and Hai-Hu Wen}\email{hhwen@nju.edu.cn}

\affiliation{Center for Superconducting Physics and
Materials, National Laboratory of Solid State Microstructures and
Department of Physics, National Collaborative Innovation Center of Advanced Microstructures, Nanjing University, Nanjing 210093, China}

\begin{abstract}
We have carried out in-plane resistivity measurements under a uniaxial pressure in NaFe$_{1-x}$Co$_x$As single crystals. A clear distinction of the in-plane resistivity $\rho_a$ and $\rho_b$ with the uniaxial pressure along $b$-axis was discovered in the parent and underdoped regime with the doping level up to about x=0.025$\pm$0.002. From the deviating point of $\rho_a$ and $\rho_b$, and the unique kinky structure of resistivity together with the published data we determined the temperatures for the nematic, structural and antiferromagnetic transitions. It is clearly shown that the nematic electronic state vanishes simultaneously with the structural transition. The antiferromagnetic state disappears however at a lower doping level. Our results, in combination with the data in BaFe$_{2-x}$Co$_x$As$_2$, indicate a close relationship between nematicity and superconductivity.

\end{abstract}

\pacs{74.70.Xa, 74.25.F-, 72.15.-v, 74.25.Dw}

\maketitle

Emergent novel states are quite popular in the correlated electronic systems. One interesting trend is that these states may replace the Fermi liquid as the normal states of unconventional superconductivity. In cuprate superconductors, for example, it has been widely known that the ``normal'' state is not normal at all. Many new electronic states, such as the stripe phase\cite{Zaanen,Tranquada,Kievelson}, checkerboard phase\cite{Hanaguri,HoffmanDavis}, possible charge-density-wave (CDW)\cite{CDWCuprate1,CDWCuprate2} etc., are discovered in the ``normal'' state of cuprate superconductors. The similar situation occurs in the iron based superconductors. In the CaFe$_{2-x}$Co$_x$As$_2$ system, an electronic state with the C$_2$ symmetry has first been discovered in the scanning tunneling measurements\cite{DavisScience}. This interesting state was later proved directly by the in-plane resistive measurements in BaFe$_{2-x}$Co$_x$As$_2$ (Ba122) with a uniaxial pressure along one of the principal in-plane axes, and thus named as the nematic phase\cite{Fisher1,Fisher2,Prozorov1}. This nematic state has also been probed by the torque measurements\cite{TorqueMatsuda} in BaFe$_2$As$_{2-x}$P$_x$, and point contact tunneling measurements in Fe$_{1-y}$Te\cite{LGreene}. It has been mentioned that this nematic state may be strongly related to the local impurities\cite{Uchida,DavisNatPhys}. Nematic states have also been revealed by the scanning tunneling spectroscopy (STS) measurements in NaFe$_{1-x}$Co$_x$As\cite{Pasupathy,WangYY}. Theoretically there are some models to interpret the nematic behavior\cite{Fernandes,Eremin,WeiKu}. It remains however to be resolved how this nematicity correlates with the structural, magnetic transitions and orbital fluctuations. In the pioneer work showing nematicity in transport measurements\cite{Fisher1,Prozorov1} on the BaFe$_{2-x}$Co$_x$As$_2$ system, the temperature difference between the structural and the antiferromagnetic (AFM) state is tiny, therefore it is difficult to resolve the issue. In present work, we have carefully measured the in-plane resistivity along a-axis and b-axis under a uniaxial pressure in the NaFe$_{1-x}$Co$_x$As system in which the structural and the AFM transitions are clearly separated. Our results shed new light in understanding the correlations among nematic, structural and AFM transitions.

\begin{figure}
\includegraphics[width=9cm]{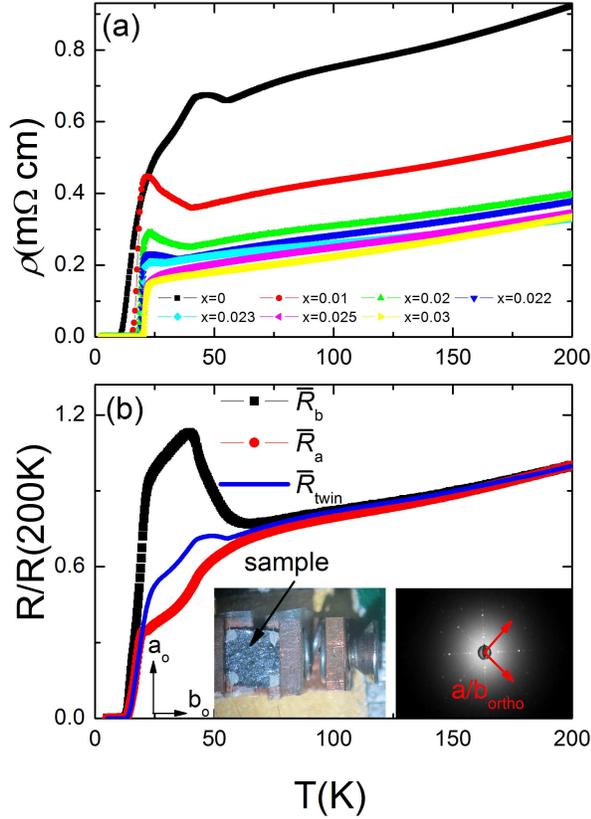}
\caption{(Color online) (a) Temperature dependence of the in-plane resistivity for twinned NaFe$_{1-x}$Co$_x$As single crystals. (b) Temperature dependence of the normalized in-plane resistance of the parent phase NaFeAs. Here $\bar{R}_a$ and $\bar{R}_b$ represent the normalized in-plane resistance along the orthorhombic $a-$ and  $b-$axis measured using the Montgomery method. The $\bar{R}_{twin}$ is measured on the same sample without uniaxial pressure. The insets show the homemade detwinning device (left) and the Laue x-ray diffraction pattern for one sample (right). }
\label{fig1}
\end{figure}

The NaFe$_{1-x}$Co$_x$As single crystals were grown by flux method using NaAs as the flux. The precursor NaAs, Fe and Co powders (both with purity 99.9\%, Alfa Aesar) were mixed together and put into an alumina crucible for growing. For details of the synthesis, one can find in our previous work\cite{Deng}. Fig. 1(a) shows the temperature dependence of the in-plane resistivity for twinned NaFe$_{1-x}$Co$_x$As single crystals. The general behavior is quite consistent with a previous study\cite{Wang}. In the underdoped region, the low-temperature upturn is related to the structural and AFM transitions (\emph{T$_s$} and \emph{T$_{AF}$}). As one can see in Fig. 1(a), with the increase of Co doping level, the structure and AFM transitions are suppressed rapidly. Although it is hard to determine the real doping level of samples for such a slight concentration difference, the systematic evolution of \emph{T$_c$}, \emph{T$_s$} and \emph{T$_{AF}$} versus the nominal doping level indicate that Co were doped into the crystal lattice successfully. Counting the very low doping level of Co, we judge that the real Co composition is close to the nominal one. Therefore we use the nominal compositions of Co in the formula. The exploration of the in-plane electronic anisotropy is difficult because it naturally forms twin boundaries in the orthorhombic phase. In order to resolve this issue, we developed a device to detwin single crystals in situ, as shown in the inset of Fig. 1(b). Laue x-ray diffraction experiments were performed in this study to confirm the orientation of the crystal axes for each crystal, as shown in the inset of Fig. 1(b). The single crystal is cut into a square shape with the edge of the sample parallel to the orthorhombic a/b-axes. In the orthorhombic phase, the shorter axes is naturally aligned in the direction of the applied strain. Similar detwinning techniques were widely used in probing the in-plane electronic anisotropy\cite{Fisher1,Ying1,Ying2,Kim,Jiang,Dhital,PCDai,XHChen}. For a review on the nematic electronic state detected by this technique, one can see reference\cite{Fisher3}. In this study, the resistivity measurements were performed on a Quantum Design instrument (PPMS-16T) using the Montgomery method\cite{Montgomery}. Compared with the usual four-probe method, the advantage of Montgomery method is that the resistance along $a-$ and $b-$axis can be measured under the same condition.

\begin{figure}
\includegraphics[width=9cm]{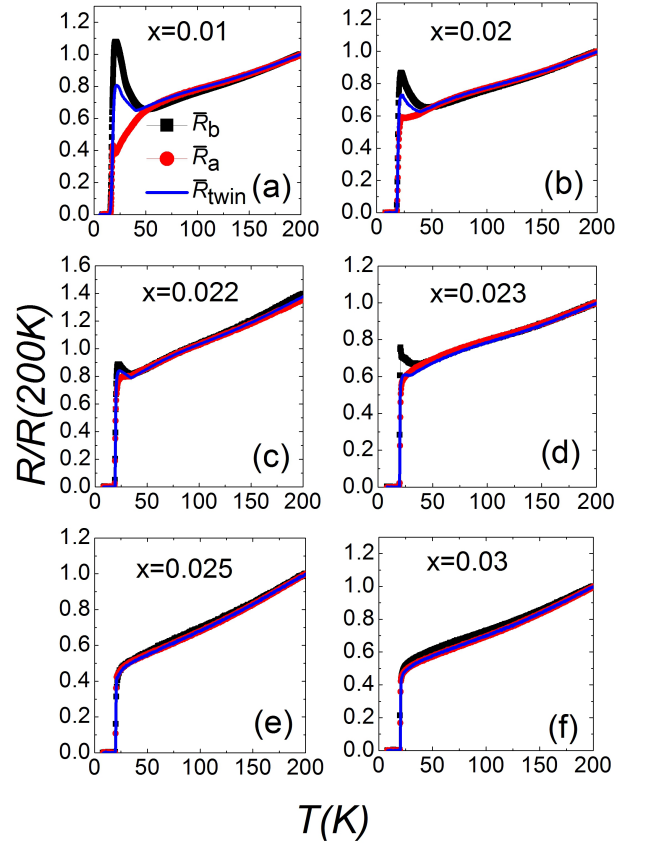}
\caption{(Color online) (a)-(f) Temperature dependence of the normalized in-plane resistance for the NaFe$_{1-x}$Co$_x$As single crystals (x=0.01, 0.02, 0.022, 0.023, 0.025, 0.03). $\bar{R}_a$ and $\bar{R}_b$ represent the normalized in-plane resistance along the orthorhombic $a-$ and orthorhombic $b-$axis measured using the Montgomery method. $\bar{R}_{twin}$ is measured on the same sample without uniaxial pressure.} \label{fig2}
\end{figure}

The temperature dependence of the normalized in-plane resistance along $a-$ and $b-$axis ($\bar{R}_a$ and $\bar{R}_b$) of the detwinned parent compound NaFeAs are shown in Fig. 1(b). For the Co-doped samples (x=0.01, 0.02, 0.022, 0.023, 0.025, 0.03), the results are shown in Fig. 2. To make a comparison of the resistivity along $a-$ and $b-$axis, the data were normalized by that measured at 200 K, except for the sample with x=0.022 (normalized at 90 K). Here $\bar{R}_{twin}$ represents the normalized in-plane resistance obtained on the same sample after releasing the uniaxial pressure. The resistivity anisotropy in the low temperature region for the underdoped samples is quite clear. The phenomenon $\bar{R}_b$ $>$ $\bar{R}_a$ in the orthorhombic phase is similar to those observed in some 122-type iron-based superconductors\cite{Fisher1,Uchida,Ying1,Ying2,Prozorov1,Prozorov2}. As shown in Fig. 1(b) and Fig. 2, for the samples with x=0, 0.01, 0.02, 0.022, 0.023, the high temperature resistivity along both $a-$ and $b-$axis decrease with temperature in a similar way, well following the trend of resistivity measured on twinned samples. With further decrease of temperature, the resistivity along $a-$axis continues to decrease, while the resistivity along b-axis starts to increase. The temperature at which $\bar{R}_a$ and $\bar{R}_b$ start to deviate from each other is defined as \emph{T$_{nem}$} (with a criterion defined below). Comparing with the measurements in Co-doped Ba122 system\cite{Fisher1,Fisher2}, we can see a difference here. The resistance along $a-$axis under a uniaxial pressure shows a progressively dropping down below \emph{T$_{nem}$}, not as a linear and smooth dropping down in Co-doped Ba122 system. It is clear that \emph{T$_{nem}$} is well above \emph{T$_s$}, indicating the existence of nematicity above the structural transition, perhaps due to the fluctuation of nematicity. With further decrease of temperature, the resistivity anisotropy $\bar{R}_b$/$\bar{R}_a$ undergoes a dramatic increase until the emergence of superconductivity. For the parent compound NaFeAs, the resistivity anisotropy $\bar{R}_b$/$\bar{R}_a$ can reach as large as 2.8. For the samples with x=0.01, the slight upturn of $\bar{R}_a$ at very low temperature is probably because the sample is not sufficiently detwinned.  The nematic, structural, antiferromagnetic temperatures (\emph{T$_{nem}$}, \emph{T$_s$} ,\emph{T$_{AF}$}) and the resistivity anisotropy in the orthorhombic phase are found to decrease monotonically with the increase of doping concentration x.  For samples with x=0.025 and 0.03, the resistivity anisotropy becomes negligible, which clearly indicates that the nematicity vanishes at around x=0.025$\pm$0.002.

\begin{figure}
\includegraphics[width=9cm]{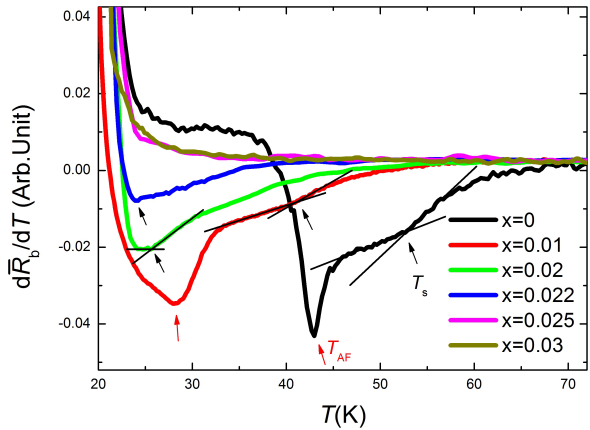}
\caption{(Color online) Derivative of the normalized resistance $\bar{R}_b$ for all samples investigated here. The arrows indicate the positions where we determine the \emph{T$_s$} and \emph{T$_{AF}$}.
} \label{fig3}
\end{figure}

Fig. 3 shows the derivative curve of the normalized resistance $\bar{R}_b$ versus T. Some kinky structures on the raw data now become more clear in the derivative curves, which allows us to determine \emph{T$_s$} and \emph{T$_{AF}$}. The \emph{T$_s$} is determined on the shoulder of the derivative curve, and \emph{T$_{AF}$} is determined at the peak position. This method has been used in previous report\cite{Wang}, and our results are well consistent with theirs. The values of $T_s$ and $T_{AF}$ are also close to those determined through the non-resistive methods, as addressed below.

\begin{figure}
\includegraphics[width=9cm]{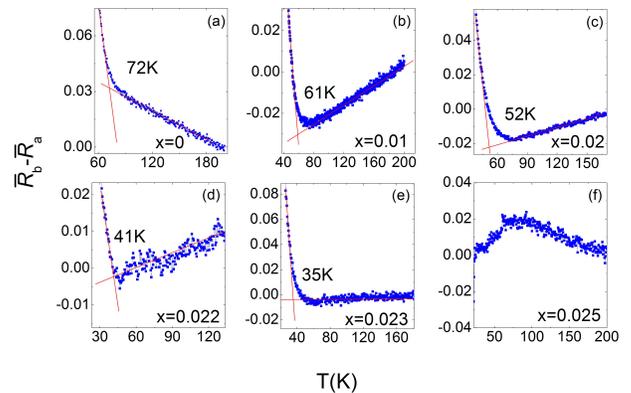}
\caption{(Color online) Difference between the normalized resistance along $b-$axis and $a-$axis. The temperature associated with the nematcicty (\emph{T$_{nem}$}) is determined using the crossing point of the normal state background line and the extrapolated linear line of the steep transition where $\bar{R}_b$-$\bar{R}_a$ changes dramatically. } \label{fig4}
\end{figure}

The temperature associated with the nematicity, named as \emph{T$_{nem}$}, is determined on the curve as shown in Fig. 4. For clarity, the temperature dependence of the difference between the normalized resistance along $b-$axis and $a-$axis is enlarged to a proper range. Here the \emph{T$_{nem}$} is determined using the crossing point of the high temperature background line and the extrapolated linear line of the steep transition where $\bar{R}_b$-$\bar{R}_a$ changes dramatically. In this way, \emph{T$_{nem}$} for the parent compound NaFeAs is estimated to be 72 K, about 20 K above the structure transition temperature. For the sample with x $\geq$ 0.025, no dramatic change can be observed on the difference curve, indicating the absence of nematicity at this doping.

In Fig. 5, we show a phase diagram containing the nematicity (\emph{T$_{nem}$}), structural (\emph{T$_s$}) and antiferromagnetic (\emph{T$_{AF}$}) temperatures combined with the \emph{T$_s$} and \emph{T$_{AF}$} data obtained from other measurement techniques\cite{Parker,Wright,Ma}. The considerable difference between the structural and the AFM state in NaFe$_{1-x}$Co$_x$As system make it possible for us to investigate whether nematicity is associated with the anti-ferromagnetic order or the structural transition or both. As one can see, the values of \emph{T$_s$} and \emph{T$_{AF}$} acquired in this study are consistent well with the data obtained from synchrotron x-ray powder diffraction (XRPD) measurements and muon spin rotation ($\mu$SR) measurements\cite{Wright}, respectively. Our data about $T_{AF}$ here are consistent very well with the $\mu$SR measurements\cite{Wright} in the same system, which gives a trend, as shown by the red dashed line, that the AFM order may vanish before x=0.02. Moreover, the \emph{T$_{AF}$} determined by a recent NMR study is roughly consistent with this trend except for the sample with x=0.0175, this discrepancy may result from the uncertainty of doping concentration in that work\cite{Ma}. Interestingly, the authors in ref.\cite{Ma} spent intensive efforts to argue that the AFM order revealed by the NMR study is completely suppressed for sample x = 0.019. On the other hand, high-resolution neutron powder diffraction (HRPD) measurements down to about 3 K reveled that the structural distortion vanishes at about x=0.025, as shown by the filled star symbol in Fig. 5. In such case, our results clearly show that the nematicity is more closely related to the structural transition, since the transition temperatures \emph{T$_{nem}$} and \emph{T$_s$} vanish simultaneously at a doping level of about x=0.025$\pm$0.002.

\begin{figure}
\includegraphics[width=9cm]{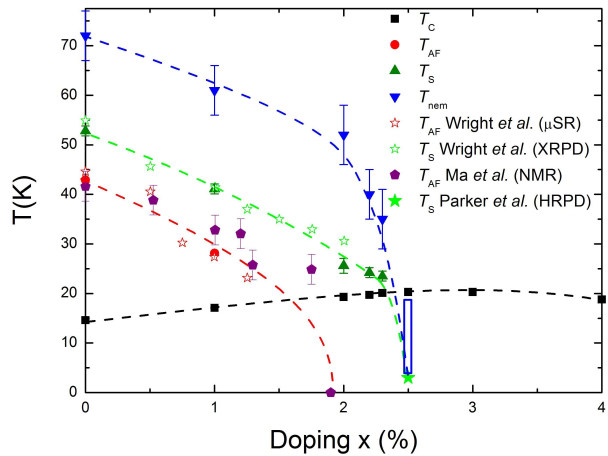}
\caption{(Color online) Phase diagram containing the nematicity (\emph{T$_{nem}$}), structural (\emph{T$_s$}) and antiferromagnetic (\emph{T$_{AF}$}) temperatures. The data obtained from other measurement techniques are collected and shown in the figure as well. For the sample with x=0.025, as highlighted by the vertical frame, we did not see any feature of structural transition and nematicity in the normal state. The HRPD measurements give one point for the structural transition at about 3 K at the doping level of x=0.025. Dashed lines are guides to the eyes. } \label{fig5}
\end{figure}

Concerning the origin of the nematic electronic state, it remains to be highly controversial. There are at least three major pictures to interpret the nematicity\cite{Fernandes}: (1) The nematicity is induced purely by the structural distortion. As soon as the lattice structure turns from the tetragonal into the orthorhombic phase, there is a distinction of the lattice constants $a$ and $b$ ($a>b$ with the strain along $b$-axis). The Landau free energy with this distortion can be easily written down\cite{Fernandes} and the lower energy state may be found with the emerging nematic order parameter.  (2) The nematicity may be related to the striped AFM state. This seems to be straightforward since the itinerant electrons detected by the transport technique are inevitable to couple with the local magnetism. Recently, an Ising like spin excitation with C$_2$ symmetry\cite{DaiPC} was found in BaFe$_{2-x}$Ni$_x$As$_2$ in associating with the in-plane anisotropy of resistivity, in supporting this picture. This picture is further supported by the Raman\cite{Yann} and optical\cite{Dusza,Nakajima} experiments. (3) The nematicity is induced by the orbital fluctuations, either due to the electronic (lifting of the degeneracy of $d_{xz}$ and $d_{yz}$) orbitals or the lattice reason.  Our data clearly show that the anisotropy of resistivity occurs at temperatures far above $T_s$ and $T_{AF}$. Other measurements, such as the tunneling on NaFe$_{1-x}$Co$_x$As also give the same observation\cite{Pasupathy,WangYY}. Our data here give an interesting hint that the nematicity and the structural transition, although not occurring at the same temperature, but vanish at the same doping level $x=0.025\pm0.002$. The AFM order disappears clearly at a lower doping level. Another interesting observation is that, for the undoped sample, the distinction $\bar{R}_b$-$\bar{R}_a$ starts to increase below $T_{nem}$, gets enhanced quickly below the structural transition $T_s$, but keeps rather stable below $T_{AF}$ since $\bar{R}_b$ drops down in parallel with $\bar{R}_a$. Therefore we would not believe that the AFM order is the right reason for the nematicity. However, we cannot rule out the possibility of magnetic fluctuations as the the driving force for the nematicity, because they do appear above $T_{AF}$. Finally we comment on the third picture, that is the electronic driven orbital fluctuations. It is well-known that the degeneracy of $d_{xz}$ and $d_{yz}$ orbitals is lifted below $T_{AF}$, as detected by the angle resolved photo-emission spectroscopy (ARPES)\cite{Yi}. Lifting of the orbital degeneracy and the striped AF magnetism seem to cast the "chicken and egg" problem. They  may entangle each other and both contribute to the formation of nematicity.

Finally, we want to point out that the nematic electronic properties seem closely related to superconductivity. Recall the data from the BaFe$_{2-x}$Co$_x$As$_2$ system, the strongest nematicity occurs at an underdoping level where superconductivity has already appeared\cite{Fisher1}. For the non-superconductive parent phase, the nematicity is much weaker. In the present NaFe$_{1-x}$Co$_x$As system, since the parent phase NaFeAs is already superconductive, we observed the strongest distinction of $\rho_a$ and $\rho_b$ from the undoped parent phase. This interesting observation is understandable by a recent argument\cite{Fernandes2} that the nematic fluctuations favor intraband pairing, which would give a boost to the s$\pm$ pairing. The argument cannot, however, be extended to the hole doped side, since in hole-doped Ba$_{1-x}$K$_x$Fe$_2$As$_2$, the in-plane resistance anisotropy $\varrho$=$\rho_b/\rho_a$-1 is very small and even changes sign compared with the electron doped Ba122 samples\cite{Ruslan}. Therefore a concise picture for describing the general connection between nematicity and superconductivity is still lacking. The superconductivity may still be induced by the magnetic fluctuations\cite{MazinPRL2008,Kuroki2008} as revealed by the interesting sign-reversal s$\pm$ gap functions\cite{Christianson,Hanaguri2,HHWen}.

In summary, through in-plane resistive measurements under a uniaxial pressure, we have successfully resolved the nematicity, structural and AFM temperatures $T_{nem}$, $T_s$ and $T_{AF}$, respectively in NaFe$_{1-x}$Co$_x$As. It is found that the structural and nematicity transitions vanish simultaneously at the doping level of about $x$ = 0.025$\pm$0.002, while the AFM order disappears at a lower doping level ($x\leq0.02$). We depict the phase diagram based on our in-plane resistive data together with earlier published data with non-resistive measurements. We propose that both the magnetic fluctuation and lifting the degeneracy of the $d_{xz}$ and $d_{yz}$ orbitals entangle each other and contribute to the nematicity. Our results also point to a close relationship between nematicity and superconductivity, at least in electron doped systems.

We thank Wei Ku, Igor Mazin, Makariy Tanatar for helpful discussions and suggestions. We appreciate the kind help from Lei Shan and Xinye Lu in establishing the uniaxial measurement setup. This work was supported by NSF of China, the Ministry of Science and Technology of China (973 projects:
2011CBA00102, 2012CB821403) and PAPD.

\end{document}